\documentclass[twocolumn,prb,showpacs]{revtex4}
\usepackage{graphicx}
\usepackage{epsfig}
\usepackage{dcolumn}
\usepackage{bm}

\begin{document} 
\title{Calculation of nanowire thermal conductivity using complete phonon dispersion
relations}
\author{N. Mingo}
\affiliation{Eloret Corp., NASA-Ames res. ctr., Moffett Field, CA 94035}
\begin{abstract}                
The lattice thermal conductivity of crystalline Si nanowires is calculated.
The calculation uses complete phonon dispersions, and does not require any
externally imposed frequency cutoffs. No adjustment to nanowire thermal conductivity measurements
is required. Good agreement with experimental results for nanowires wider than 35 nm
is obtained. A formulation in terms of the transmission function is given.
Also, the use of a simpler, nondispersive "Callaway formula", is discussed from the 
complete dispersions perspective.
\end{abstract}

\pacs{63.22.+m,65.80.+n,66.70.+f,63.20}
\maketitle

\begin{text}
Determining the thermal conductivity of semiconducting nanowires plays a crucial role in the 
development of a new generation of thermoelectric materials \cite{Mahan}. However, experimental results
are still scarce \cite{Fon}.
The first measurements of the thermal
conductivity of Si nanowires in the $\sim 20-115 nm$ range have been recently reported in ref.\onlinecite{Li}.
From a theoretical point of view, it is important to be able to quantitatively calculate lattice
thermal conductivities of nanowires in a predictive fashion. In the past,
several techniques have become widespread in the calculation of 
thermal conductivities of bulk materials \cite{Callaway,Holland,Parrott,Berman}. These works are based on
{\it linearized dispersion} models that involve
a certain number of adjustable parameters. 
In principle, one might hope that a proper account of the boundary scattering
in the nanowires might enable us to predict nanowire thermal conductivities
using the procedures of ref.~\onlinecite{Callaway} or ref.~\onlinecite{Holland}, without any
parameters other than those obtained from fitting the bulk data. 
Nevertheless, calculations using those two approaches did not yield 
reasonable results for nanowires \cite{APL}. 
A first goal of this paper is to show that, using accurate phonon dispersion relations,
{\it predictive} calculations of the nanowire thermal conductivities can be done,
which {\it do not involve any adjustable parameter to fit the nanowire experimental curves}.
A second result in this paper shows that the use of a "modified" Callaway approach is still
possible for nanowires, although this second approach is not predictive, and requires adjustment
to nanowire measurements.

For a nanowire suspended between two thermal reservoirs,
its thermal conductance, $\sigma(T)$, is defined as the flow of heat current through the wire
when a small temperature difference $\Delta T$ exists between the reservoirs,
divided by $\Delta T$ and taking the limit $\Delta T \rightarrow 0$. If the
phonons transmit ballistically, the conductance is
\begin{eqnarray}
\sigma_{b}(T)=\sum_{\alpha}\int_0^{\pi/a_z} {dk_z\over 2\pi}[\hbar \omega v_z(\alpha,k_z)
{df_{B}\over dT}]
\end{eqnarray}
where $\alpha$ is a set of discrete quantum numbers labeling the 
particular subbands in the one dimensional phonon dispersion relations.
$f_B$ is the Bose distribution, and $v_z$ is the speed of the phonon in the axial
direction of the nanowire, $z$.
Using $v_z dk_z=d\omega$ we have \cite{Rego}
\begin{eqnarray}
\sigma_{b}(T)=\sum_{\alpha}\int_{\omega^i_{\alpha}}^{\omega^f_{\alpha}}
{d\omega\over 2\pi}[\hbar \omega {df_{B}\over dT}]
\equiv \int_0^{\infty} \Xi_{b}(\omega){\hbar \omega \over 2\pi}{df_{B}\over dT}d\omega
\label{Eq:conductance}
\end{eqnarray}
where $\omega^{i(f)}_{\alpha}$ is the lower (upper) frequency limit of subband
$\alpha$. Here we have defined the ballistic transmission function, $\Xi_b(\omega)$,
which simply corresponds to the
number of phonon subbands crossing frequency $\omega$.

However, in general, phonon scattering takes place in the wire and at the 
contacts, so that the transmission function is not given by the simple mode
count stated above. In general, the transmission function can be obtained from
the Green function of the system. This allows to study
the interesting issues of ballistic versus diffusive transport \cite{deJong},
or the effect of one dimensional localization.\cite{Beenakker}
We shall not be concerned with those phenomena here, but shall assume that
diffusive transport takes place and the Boltzmann transport 
equation is valid. From this standpoint, it 
has been shown that a mode's transmission can be described in terms of
a characteristic relaxation length in the wire's direction, 
$\lambda_{\alpha}(\omega)$, such that
$\sigma_{diff}(T)\simeq \sum_{\alpha}\int {\lambda_{\alpha}(k_z)\over L} 
{\hbar \omega_{\alpha}(k_z) \over 2\pi}{df_{B}\over dT}v_z(\alpha,k_z)d k_z$
where $L$ is the nanowire's length \cite{Klemens}. 
The thermal conductivity is defined as $\kappa(T)={L\over s}\sigma$, where $s$
is the cross section of the nanowire. Therefore, in this diffusive regime
\begin{eqnarray}
\kappa(T)={1\over s}\sum_{\alpha} \int_0^{\pi/a_z} \lambda_{\alpha}(k_z)
{\hbar \omega_{\alpha}(k_z) \over 2\pi}{df_{B}\over dT}v_z(\alpha,k_z)d k_z.
\end{eqnarray}
To obtain the thermal conductivity we must compute the complete dispersion
relations for the wire, $\omega_{\alpha}(k_z)$, which, using 
$v_z(\alpha,k_z)={d\omega_{\alpha}(k_z)\over dk_z}$, allows us to express
the previous equation as 
\begin{eqnarray}
\label{Eq:thcond1}
\kappa(T)={1\over s}\int_0^{\infty} (\sum_{\alpha} \lambda_{\alpha}(\omega))
{\hbar \omega \over 2\pi}{df_{B}\over dT}d\omega,\\
\label{Eq:lambdas}
\hbox{with }\ \lambda_{\alpha}(\omega) \equiv \left\{ \matrix{
\lambda_{\alpha}(k_z(\alpha,\omega)), & \omega_{\alpha}^i < \omega < \omega_{\alpha}^f \cr
0, & \hbox{otherwise.} \cr} \right.
\end{eqnarray}

If the modes' lifetimes $\tau(\omega)$ 
are independent of the subband, and only depend on the frequency, then,
the relaxation lengths are given by
\begin{eqnarray}
\lambda_{\alpha}(\omega)=v_z(\alpha,\omega)\tau(\omega),&& \omega_{\alpha}^i < \omega < \omega_{\alpha}^f 
\label{Eq:lambdas2}
\end{eqnarray}
Knowing $\omega_{\alpha}(k_z)$ and $\tau(\omega)$, Eqs.~(\ref{Eq:thcond1}-
\ref{Eq:lambdas2}) already
allow us to compute the thermal conductivity.
Nevertheless, it is useful to recast Eq.~(\ref{Eq:thcond1}) into an equivalent form, so as
to explicitly retain the "transmission function" concept. For this, we define $N_b(\omega)$
as the number of phonon subbands crossing frequency $\omega$: \footnote{$N_b$ is equivalent to the 
"ballistic" transmission function.}
\begin{eqnarray}
\label{Eq:N}
N_b(\omega) \equiv \sum_{\alpha}\Theta(\omega-\omega_{\alpha}^i)\Theta(\omega_{\alpha}^f-\omega),\\
\Theta(x) = \left\{ \matrix{
1 & x \ge 0 \cr
0 & x < 0. \cr} \right.
\end{eqnarray}
The average group velocity in the axial direction is defined as
\begin{eqnarray}
\langle v_z(\omega)\rangle \equiv \left (\sum_{\alpha} v_z(\alpha,\omega)\right )/N_b(\omega)
\label{Eq:vave}
\end{eqnarray}
where the sum extends to subbands $\alpha$ such that $\omega_{\alpha}^i<\omega<\omega_{\alpha}^f$.
From Eqs.~(\ref{Eq:lambdas}-\ref{Eq:lambdas2}) and (\ref{Eq:vave}), it follows that $\sum_{\alpha}\lambda_{\alpha}(\omega) = \tau(\omega)\sum_{\alpha}
v_z(\alpha,\omega) = \tau(\omega)N_b(\omega)\langle v_z (\omega) \rangle$. Thus Eq.~(\ref{Eq:thcond1})
can be recast into its completely equivalent form \footnote{Note the similarity with Eq.~(\ref{Eq:conductance}).}
\begin{eqnarray}
\label{Eq:thcond2}
\kappa(T)={L\over s}\int_0^{\infty} \Xi(\omega) {\hbar \omega \over 2\pi}{df_{B}\over dT}d\omega\\
\hbox{with } \ \ {L\over s} \Xi(\omega)= \tilde N_b(\omega) \tau(\omega) \langle v_z(\omega)\rangle.
\label{Eq:transmission}
\end{eqnarray}
Here, the diffusive transmission function $\Xi$ has been defined, and
\begin{eqnarray}
\tilde N_b(\omega)\equiv N_b(\omega)/s.
\end{eqnarray}
Computation of the transmission function requires the obtention of the full phonon dispersions
for the nanowire.
We now proceed to separately study
each of the three factors in Eq.~(\ref{Eq:transmission}):
$\tilde N_b(\omega)$, $\tau(\omega)$, and $\langle v_z(\omega)\rangle$.

$\tilde N_b(\omega)$ is obtained from the complete dispersion relations for the wire. To compute
them, we have used the 
interatomic potential proposed by ref.~\onlinecite{Harrison}, in which the energy of the system is
given as the sum of two and three body terms. The two body terms are defined as
$$\delta E_0 (i,j)= \hbox{$ {1\over 2} $ } C_0 {(d_{i,j} - d_0)^2 / d_0^2}$$
for every pair of nearest neighbors $i$ and $j$, where $d_{i,j}$ is the distance between the atoms and
$d_0$ is the lattice equilibrium distance. The three body terms are defined as
$$\delta E_1(i,j,k)=\hbox{$ {1\over 2} $ } C_1 \delta \Theta_{i,j,k}^2$$
for every pair of bonds joining atoms $i$, $j$ and $k$, 
where $\delta\Theta$ is the deviation with respect to the
equilibrium angle between the two bonds in the lattice. 
The constants are $C_0=49.1eV$ and $C_1=1.07eV$ \cite{Harrison}.
With this potential the
dynamical matrix of the system is constructed, and the dispersion relations 
for the wire are calculated. The inset of fig.~\ref{Fig:trans} shows one
example. The bulk dispersions obtained with the Harrison potential are close
to those experimentally measured. Although a more sophisticated calculation of the
interatomic potential is possible using ab initio techniques, this would only result
in minor differences in the calculated $\tilde N_b$ and the conductivities computed from it.

The dispersions in general depend on the cross sectional shape of the wire, and 
on whether its surface is reconstructed or not, clean or coated by an overlayer, etc.
As the surface/volume ratio decreases, these differences become smaller, and for wide
enough wires of cross section $s$, the function $\tilde N_b(\omega)\equiv N_b(\omega)/s$
approaches a limiting form, $\tilde N_b^{\infty}(\omega)$, independent
of the surface features.
To ascertain whether it is valid to approximate $\tilde N_b(\omega)$ by 
$\tilde N_b^{\infty}(\omega)$ for nanowires wider than 35 nm, the former was
calculated for nanowires of increasing width, with frozen boundary conditions. 
(As compared to "free boundary", frozen boundary conditions are more similar to the experimental 
situation, where 1 nm of $SiO_2$ coats the wire.)
For a romboidal cross section wire of 17 nm side, $\tilde N_b$ deviates from the
limiting value $\tilde N_b^{\infty}$ by less than $0.1 \%$.
\footnote{$\tilde N_b$ was also calculated for a "free boundary" 
$11\times 11$ unit cells cross section wire, verifying that
it is already quite similar to $\tilde N_b^{\infty}$, and very similar results for $\kappa$
were yielded when $\tilde N_b$ was replaced by $\tilde N_b^{\infty}$,
keeping the remaining parameters unchanged.} Therefore, for the widths considered here
($l>35 nm$), $\tilde N_b=\tilde N_b^{\infty}$ is a valid assumption. In the thermal conductivity
calculations presented here, $\tilde N_b^{\infty}$ for the 110 direction was used (fig.~\ref{Fig:trans}).

\begin{figure}
\includegraphics[width=8. cm]{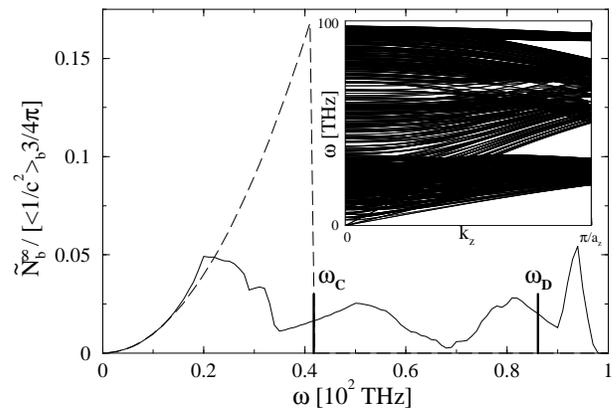}
\caption{{\it Solid line:} 
function $\tilde N_b^{\infty}(\omega)$ for the (110) direction, calculated using
complete phonon dispersions. {\it Dotted line:} parabolic approximation to $\tilde N_b^{\infty}(\omega)$, 
for a non dispersive medium. {\it Inset:} dispersion relations for a 2.2 nm wide wire with frozen boundary.
}

\label{Fig:trans}
\end{figure}

The phonon lifetime is commonly given by Matheissen's rule, expressing the
total inverse lifetime as the sum of the inverse lifetimes corresponding
to each scattering mechanism.
For bulk or a macroscopically
thick whisker, the expressions used for boundary, anharmonic and impurity scattering are \cite{Asen-Palmer}:
\begin{eqnarray}
\tau^{-1}_b =& (\langle 1/c \rangle_b lF)^{-1}\nonumber\\
\label{Eq:times}
\tau_a^{-1} =& B T \omega^2 e^{-C/T} \\
\tau^{-1}_i =& A \omega^4\nonumber
\end{eqnarray}
From here on, $\langle \rangle_b$ denotes
averaging in the three acoustic branches, i.e. 
$\langle 1/c \rangle_b \equiv {1\over 3}\sum_{i=1}^3 1/c_i$,
where $c_i$ is the speed of sound for each branch. (Do not confuse with 
averages over subbands, denoted by $\langle\rangle$.) A, B and C are numerical
constants specified later.
$l$ is the wire's lateral dimension, and
$F$ relates the boundary scattering rate to the shape and specularity of the
sample's boundary. 
Eqs.~(\ref{Eq:times}) are always valid for wire thicknesses larger than a certain threshold, $L_{3D}$.
On the other hand, if the width of the whisker is very small, such that $L<<L_{3D}$,
the frequency difference between consecutive phonon subbands gets large. This confinement
modifies the inter-subband scattering, and the validity of Eqs.~(\ref{Eq:times})
breaks down. \footnote{The case of strong confinement can be dealt with by 
using modified expressions for $\tau$ involving the subbands' group velocities
rather than the speed of sound, as done in ref.~\onlinecite{Zou}.}
A criterion to estimate $L_{3D}$ is: confinement is important if the energy spacings
between consecutive subbands, $\hbar \Delta\omega$, are larger or comparable to the
thermal energy $k_B T$. The largest intersubband spacings are of the order of 
$\hbar\Delta\omega\sim {2\pi\over l}\hbar c$. Thus, for Eqs.~\ref{Eq:times} to be valid at
temperatures $40^oK$ or higher, we need $l>10 nm$. This is well fulfilled for nanowires
in the 37-115 nm range, as the ones experimentally measured \cite{Li}. Therefore,
for these widths, we use Eqs.~\ref{Eq:times} without modification. 
The good agreement with experimental results corroborates their validity.
For sub 20 nm widths, confinement effects may be expected.
\footnote{Measurements of narrow nanowires around $20 nm$ wide (not shown here) have yielded
thermal conductivities much smaller than those predicted,
suggesting differences in the transport process at those small diameters.}

Anharmonic scattering should include the effects of umklapp as well as
normal processes. In some works normal processes have been omitted in favor of 
umklapp scattering, without extensive discussion \cite{Fon,Balandin}. In general,
if resistive processes dominate, the normal relaxation time can be disregarded 
\cite{Klemens}. For the higher frequencies, umklapp scattering is proportional to
$\omega^2$, and it thus dominates over normal processes, which depend linearly on
frequency \cite{HanKlemens}. At the lower frequencies, in the case of nanowires,
boundary scattering dominates. This justifies the form of $\tau_a$ in Eq.~\ref{Eq:times}, which is
the functional dependence for umklapp scattering at the higher frequencies.
Constants B and C were adjusted to reproduce the {\it bulk material}'s experimental thermal
conductivity curve, yielding $B=1.73\times 10^{-19} s/K$ and $C=137.3K$. 
\footnote{An estimation of the order of magnitude of B can be made as \cite{Zou}
$B\sim 2\gamma^2 {k_B \over \mu V_0 \omega_D}\sim 10^{-19}$,
which agrees with our obtained value 
($\gamma=$ Gruneisen parameter, $\mu=$ shear modulus, $V_0=$ volume per
atom, and $\omega_D=$ Debye frequency.)
Constant $C$ has the order of magnitude of the transverse phonons' frequency 
near the zone edge \cite{Han,Balandin}.}
Parameter $A=1.32\times 10^{-45}s^3$ is analytically determined
from the isotope concentration, and it should not be adjusted \cite{Asen-Palmer}, so we maintain
the value given in [\onlinecite{Holland}]. 

The third term in Eq.~(\ref{Eq:transmission}), $\langle v_z(\omega)\rangle$, is also calculated from the
dispersion relations. In a frozen boundary wire, it is a matter of simple geometry to prove that
at the lower part of the spectrum,\footnote{
For this, use $\omega_{\alpha}(k_z)\simeq c_i\sqrt{k_z^2+k_{\perp}^2}$, $\alpha \equiv \{i,\vec k_{\perp}\}$, and
$\langle v_z \rangle \simeq {\sum_{i=1}^3\int_0^{\omega/c_i} c_i {\sqrt{(\omega/c_i)^2-k_{\perp}^2}
\over \omega/c_i}2\pi k_{\perp}
d k_{\perp}\over \sum_{i=1}^3\int_0^{\omega/c_i} 2\pi k_{\perp} d k_{\perp}}$.}
\begin{eqnarray}
\langle v_z \rangle \simeq 
{2\over 3}\langle 1/c \rangle_b / \langle 1/c^2 \rangle_b
\label{Eq:linspeed}
\end{eqnarray}
where 
$\langle 1/c^2 \rangle_b \equiv {1\over 3} \sum_{i=1}^3 1/c^2_i$.
The relaxation times, Eq.~(\ref{Eq:times}) are quite rough, since they only depend on the
lower spectrum speeds of sound. Thus, no additional approximation is incurred if
Eq.~(\ref{Eq:linspeed}) is used for the third term in Eq.~(\ref{Eq:transmission}).

Fig.~\ref{Fig:thcond} shows the results of the calculation using the {\it complete dispersions} 
transmission function into Eq.~(\ref{Eq:thcond2}).
Calculations for nanowires of diameters $F l = 38.85$, $72.8$
and $132.25$ nm are shown, together with experimental results from ref.~\onlinecite{Li}. 
Although the TEM measured diameters
of the experimental nanowires were 37, 56 and 115 nm, variations in cross sectional
shape, as well as thickness of the nanowire's oxide surrounding layer are possible, so that
the cross sectional area is not accurately known. Assuming $l$ equal to the TEM diameters,
values can be assigned to $F$, which in the current case are $F\simeq 1.05$, $1.3$ and $1.15$.
These values suggest that the boundary scattering is very diffusive, rather than
specular. 

\begin{figure}
\includegraphics[width=6.5 cm]{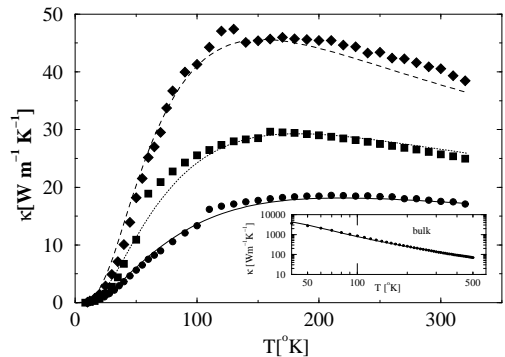}
\caption{Thermal conductivities versus temperature calculated using the complete
dispersions transmission function,
for $F l=1.05\times 37 nm$
(solid), $1.3\times 56 nm$ (dotted) and $1.15\times 115 nm$ (dashed).
Dots: experimental results from ref.~\onlinecite{Li}. {\it Inset:} bulk.
}
\label{Fig:thcond}
\end{figure}

Let us see what would happen if, instead of the atomistically calculated complete dispersions,
a nondispersive medium were used. In such a case, for frozen boundary, function $\tilde N_b^{\infty}(\omega)$
is analytical, having a parabolic dependence:
\begin{eqnarray}
\hbox{(non dispersive)}\
\  \tilde N_b^{\infty}(\omega) \simeq &\sum_{i=1}^3\int_0^{\omega/c_i} 2\pi k_{\perp} 
{dk_{\perp}\over (2\pi)^2} \nonumber \\
=&{3\over 4\pi}\omega^2 \langle 1/c^2 \rangle_b
\label{Eq:NCallaway}
\end{eqnarray}
Substituting Eq.~(\ref{Eq:NCallaway}) into Eqs.~(\ref{Eq:thcond2}) and (\ref{Eq:transmission}), 
the Callaway formula \cite{Callaway,Holland} is obtained. Since now the subbands extend to
infinite frequency, an upper cutoff has to be imposed, 
which was not necessary when using the complete dispersions, because for these ones the 
upper bands' limits are well defined. Traditionally,
the Debye frequency, 
$\omega_D$ (86~THz for Si), has been used as cutoff \cite{Callaway}. If we do this, the 
calculated thermal conductivities for nanowires are in large disagreement with experimental
measurements. As fig.~\ref{Fig:Callaway}(a) shows, no calculated curve agrees in shape with any
experimental curve: the theoretical inflection points are always too high. Comparison of the 
parabolic approximation to $\tilde N_b^{\infty}$ (Eq.~\ref{Eq:NCallaway}) with the correct $\tilde N_b^{\infty}$, in 
fig.~\ref{Fig:trans}, shows that the former grossly overestimates the transmission
of the high frequency phonons, if its cutoff is set at $\omega_D$. 
Nevertheless, if a lower cutoff $\omega_C$ is chosen, the "Callaway" and "correct" 
$\tilde N_b^{\infty}(\omega)$ curves are more similar. Using $\omega_C$, the shapes of the 
theoretical thermal conductivity curves agree well with the experimental 
ones (fig.~\ref{Fig:Callaway}(b)).
The values of the lifetime parameters B and C are those that yield the correct bulk limit:
$3.9\times 10^{-19}s/K$ and $223.1^oK$ for fig.~\ref{Fig:Callaway}(a),
and $1.7\times 10^{-19}s/K$ and $151.8^oK$ for fig.~\ref{Fig:Callaway}(b). 

For bulk, it is possible to obtain a good fit of experimental data using either $\omega_C$
or $\omega_D$ (see inset of fig.~\ref{Fig:Callaway}), because the dominant umklapp scattering 
makes the role of high frequencies
less important. For nanowires, boundary scattering dominates, and high frequencies play a more
important role. Therefore, using the experimental Debye frequency as a cutoff for the Callaway
formula does not yield correct results for nanowires, although good results can be obtained with
a lower, adjusted cutoff. This one must be determined by comparison with nanowire experimental
results ({\it non predictive} approach). On the other hand, if the atomistically calculated, 
{\it complete dispersion} relations are used, good agreement with experimental results is 
obtained (fig.~\ref{Fig:thcond}),
with no need for prior knowledge of any experimental measurements for nanowires.

\begin{figure}
\includegraphics[width=8.5 cm]{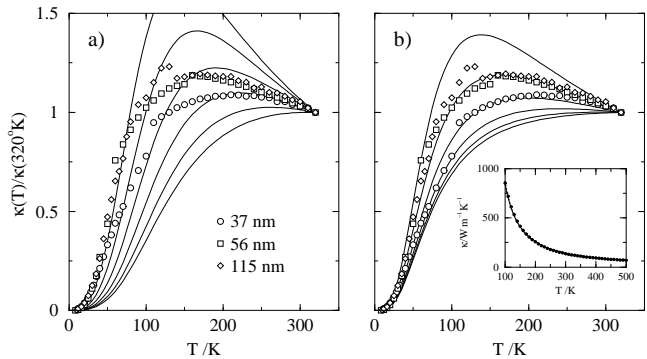}
\caption{Thermal conductivity normalized by its value at $320^oK$, for nanowires of 
different widths, calculated by the Callaway formula with cutoff at $\omega_D$ (86 THz) (a), and
at $\omega_C$ (42 THz) (b). (Experimental results from ref.~\onlinecite{Li}.) 
{\it Inset:} bulk thermal conductivity for the two cases (overlapping almost
completely) and experimental curve from ref.~\onlinecite{Holland} (dots).
}
\label{Fig:Callaway}
\end{figure}

To conclude, it was shown that, by using complete, atomistically computed phonon dispersions,
it is possible to {\it predictively} calculate lattice thermal conductivity curves for nanowires,
in good agreement with experiments. On the other hand, it was not possible to obtain 
correct results with the approximated Callaway formula. Still, good results could be 
obtained with a modified Callaway formula, although non predictively. The results of this
paper are only expected to apply to "nanowhiskers" for which phonon confinement effects are
unimportant. Si nanowires wider than $\sim 35 nm$ are within this category.

I am indebted to D. Li, A. Majumdar and Ph. Kim for communication of their experimental data, 
and discussions, and to A. Balandin and Liu Yang for discussions.

\end{text}

\end{document}